\documentclass[
    ,final            
  ]
  {aipproc}

\layoutstyle{8x11single}

\usepackage{amsmath}
\usepackage{slashed}

\newcommand{\beq}{\begin{equation}}
\newcommand{\eeq}{\end{equation}}
\newcommand{\beqa}{\begin{eqnarray}}
\newcommand{\eeqa}{\end{eqnarray}}
\newcommand{\no}{\nonumber}
\newcommand{\q}{\quad}

\newcommand{\tr}{\mbox{tr}}
\newcommand{\sfrac}[2]{{\textstyle\frac{#1}{#2}}}

\def\bra#1{\left\langle #1\right|}
\def\ket#1{\left| #1\right\rangle}

\newcommand{\newop}[2]{\def#1{\mathop{\mathrm{#2}}\nolimits}}
\newop{\artanh}{artanh}
\newop{\det}{det}
\newop{\tr}{tr}
\newop{\diag}{diag}
\newop{\Re}{Re}
\newop{\Im}{Im}

\newcommand{\Lagr}{\mathcal{L}}

\begin{document}

\title{The $\bar{K} N \rightarrow K \Xi$ reaction in coupled channel chiral models up to next-to-leading order}


\author{V.K. Magas}{
  address={Dept. d'Estructura i Constituents de la Mat\`eria, Universitat de Barcelona,\\ Mart\'i Franqu\`es 1, E08028 Barcelona, Spain}
  ,altaddress={Institut de Ci\`encies del Cosmos, Universitat de Barcelona,\\
 Mart\'i Franqu\`es 1, E08028 Barcelona, Spain}
}

\author{A. Feijoo}{
  address={Dept. d'Estructura i Constituents de la Mat\`eria, Universitat de Barcelona,\\ Mart\'i Franqu\`es 1, E08028 Barcelona, Spain}
}

\author{A. Ramos}{
  address={Dept. d'Estructura i Constituents de la Mat\`eria, Universitat de Barcelona,\\ Mart\'i Franqu\`es 1, E08028 Barcelona, Spain}
   ,altaddress={Institut de Ci\`encies del Cosmos, Universitat de Barcelona,\\
 Mart\'i Franqu\`es 1, E08028 Barcelona, Spain}
}

\begin{abstract}
We study the meson-baryon interaction in S-wave in the strangeness S=-1 sector using a chiral unitary approach 
based on a next-to-leading order chiral SU(3) Lagrangian. We fit our model to the large set of experimental  
data in different two-body channels. We pay particular attention to the $\bar{K} N \rightarrow K \Xi$ reaction, where 
the effect of the next-to-leading order terms in the Lagrangian are sufficiently large to be observed, since at tree level the cross section of this reaction is zero. 
For these channels we improve our approach by  phenomenologically taking into account effects of the high spin hyperonic resonances. 
\end{abstract}

\maketitle


\section{Introduction}

In this work we aim to improve the chiral model description of the low energy hadron dynamics  by taking into account effects of the next-to-leading (NLO) terms in the chiral Lagrangian.  It is well known that Quantum Chromodynamics (QCD), the proper theory of strong interactions, is not applicable to study low energy hadron reactions. For such studies effective theories should be used, and SU(3) Chiral Perturbation Theory  ($\chi$PT) is a classical example. This theory is based on an effective Lagrangian with hadron degrees of freedom, which respects the symmetries of QCD, in particular the chiral symmetry $SU(3)_R\times SU(3)_L$.
In spite of the many successful applications of $\chi$PT, it fails to describe the hadron dynamics in the vicinity of dynamically generated resonances. A good example of such situation is the kaon-nucleon interaction at low momenta, where the perturbation scheme is violated by the presence of the  $\Lambda(1405)$ resonance, located only $27$ MeV below the $\bar{K}N$ threshold. In this case the use of some non-perturbative techniques is mandatory. In particular such a situation can be successfully studied within a unitary extension of Chiral Perturbation Theory (U$\chi$PT), originally proposed in \cite{ref5}, where the unitarization is implemented in coupled channels.

Interestingly, the $\Lambda(1405)$ resonance does not only give a reason to use U$\chi$PT theory, but also provides a good test of the predictive power of this approach. The point is that the $\Lambda(1405)$ is a dynamically generated resonance. This was predicted  for the first time  in 1977, see Ref. \cite{L1405}, and later detailed calculations performed in the framework of U$\chi$PT have shown that $\Lambda(1405)$ is actually a superposition of two close dynamically generated states. Most of such simulations \cite{2pole,Borasoy:2006sr,ref3,ref22} predict one state at energy $\approx 1390$ MeV with larger width $\approx 130$ MeV, which couples most strongly to $\Sigma \pi$ channels; and the other one at higher energy $\approx 1420$ MeV and with a much narrower width $\approx 30$ MeV, which couples most strongly to $\bar{K} N$ channels. \footnote{Although a recent study \cite{Mai:2012dt} predicts rather different positions for the two $\Lambda(1405)$ state. This work  is also interesting in connection with our study because the authors perform calculations based on NLO Chiral Lagrangian, but they do not take into consideration the $\eta$ and $\Xi$ channels (their effects are "included" by renormalizing other model parameters), while for us the latter ones are of crucial importance. } Thus, the experimental shape of the $\Lambda(1405)$ resonance depends on the details of the given experiment, namely on the relative weight of the $\Sigma \pi$ and $\bar{K} N$ channels in the given reaction. This rather nontrivial prediction has been finally confirmed experimentally, see Ref. \cite{PRL} for more details. 

\section{Formalism}

In this section, we briefly outline the coupled-channel formalism  
of meson-baryon scattering. It is based on the SU(3) chiral 
effective Lagrangian which incorporates the same symmetries and symmetry breaking 
patterns as QCD and describes the coupling of the pseudoscalar octet
$(\pi, K, \eta)$ to the ground state baryon octet $(N, \Lambda, \Sigma, \Xi)$.
The Lagrangian
\beq \label{Lagr}
\Lagr = \Lagr_\phi + \Lagr_{\phi B}
\eeq
includes the mesonic term $\Lagr_\phi$ up to second chiral order \cite{GL},
\beq \label{Lagrphi} 
\Lagr_\phi =  \frac{f^2}{4} \langle u_{\mu} u^{\mu} \rangle 
            + \frac{f^2}{4} \langle \chi_+ \rangle , 
\eeq
where $\langle \dots \rangle$ denotes the trace in flavor space.
The pseudoscalar meson octet $\phi$ is arranged in a matrix valued field
\beq \label{Uphi}
U(\phi) = u^2(\phi) = \exp{\left( \sqrt{2} i \frac{\phi}{f} \right)} 
\eeq
and $f$ is the pseudoscalar decay constant in the chiral limit. The quantity $U$ enters the 
Lagrangian in the combinations $u_\mu = i u^\dagger \partial_\mu U u^\dagger$
and $\chi_+ = 2 B_0 (u^\dagger \mathcal{M} u^\dagger + u \mathcal{M} u)$, 
the latter one involving explicit chiral symmetry breaking via the quark mass
matrix $\mathcal{M} = \diag{(m_u, m_d, m_s)}$, and 
$B_0 = - \bra{0} \bar{q} q \ket{0} / f^2$ relates to the order parameter of 
spontaneously broken chiral symmetry.

The second piece of the Lagrangian in eq.~(\ref{Lagr}), $\Lagr_{\phi B}$,  describes the 
meson-baryon interactions and reads at lowest order \cite{K}
\beq \label{LagrphiB1}
\Lagr_{\phi B}^{(1)}  =  i \langle \bar{B} \gamma_{\mu} [D^{\mu},B] \rangle
                           - M_0 \langle \bar{B}B \rangle 
                         - \frac{1}{2} D \langle \bar{B} \gamma_{\mu}
                             \gamma_5 \{u^{\mu},B\} \rangle
                           - \frac{1}{2} F \langle \bar{B} \gamma_{\mu} 
                             \gamma_5 [u^{\mu},B] \rangle .
\eeq
The $3 \times 3$ matrix $B$ collects the ground state baryon octet, 
$M_0$ is the common baryon octet mass in the chiral limit and $D$, $F$
denote the axial vector couplings of the baryons to the mesons. Their numerical 
values can be extracted from semileptonic hyperon decays and
we employ the central values determined in \cite{CR}: $D = 0.80$, $F = 0.46$.
The covariant derivative of the baryon field is given by
\beq \label{CoDer}
[D_\mu, B] = \partial_\mu B + [ \Gamma_\mu, B]
\eeq
with the chiral connection
\beq
\label{Conn}
\Gamma_\mu = \sfrac{1}{2} [ u^\dagger,  \partial_\mu u] . 
\eeq
We also need
the next-to-leading order contribution to $\Lagr_{\phi B}$ which is given by
\beqa \label{LagrphiB2}
\Lagr_{\phi B}^{(2)} & = &   b_D \langle \bar{B} \{\chi_+,B\} \rangle
                           + b_F \langle \bar{B} [\chi_+,B] \rangle
                           + b_0 \langle \bar{B} B \rangle \langle \chi_+ \rangle \no \\ 
                     &   & + d_1 \langle \bar{B} \{u_{\mu},[u^{\mu},B]\} \rangle
                           + d_2 \langle \bar{B} [u_{\mu},[u^{\mu},B]] \rangle 
                           + d_3 \langle \bar{B} u_{\mu} \rangle \langle u^{\mu} B \rangle
                           + d_4 \langle \bar{B} B \rangle \langle u^{\mu} u_{\mu} \rangle ,
\eeqa
where only the pieces relevant for our analysis are displayed.

Then the interaction kernel for $(i,j)$ channels, $V_{ij}$, can be calculated from the chiral Lagrangian up to the corresponding order in momentum over baryon mass. For the meson-baryon interaction, the lowest order term in momentum, i.e. leading order (LO) term, is the so called Weimberg-Tomozawa (WT) term: 
\begin{equation}
V^{WT}_{ij}=-C_{ij}\frac{1}{4f^2}\bar{u}_{B'}^{s'}(p')\gamma^\mu u_B^{s}(p)(k_\mu+k'_\mu)\,, 
\label{WT}
\end{equation}
which depends only on one parameter - the pion decay constant $f$. $C_{ij}$ is a matrix of coefficients; $k^\mu$ and  $k'^\mu$ are the four-momenta for the incoming and outgoing mesons in the process; $u_B^{s}(p)$ is the incoming baryon with $s$ spin and $p$ momentum, and analogously for the $\bar{u}_{B'}^{s'}(p')$ of the outgoing baryon. The pion decay constant is well known experimentally, $f_{exp}=93.4~MeV$, however in LO U$\chi$PT calculations this parameters is usually taken to be $f=1.15-1.2 f_{exp}$, in order to partly simulate the effect of the higher order corrections.

The interaction kernel up to NLO is also known:
\begin{equation}
V^{NLO}_{ij}=V^{WT}_{ij}+\frac{1}{f^2}\bar{u}_{B'}^{s'}(p')\left(D_{ij}-2(k_{\mu}k'^{\mu})L_{ij}\right)u_B^{s}(p)\sqrt{\frac{M_i+E_i}{2M_i}}\sqrt{\frac{M_j+E_j}{2M_j}}
\label{V_NLO}
\end{equation} 
where  $D_{ij}$ and $L_{ij}$ are the coefficient matrixes, which depend on the new parameters: $b_0$, $b_D$, $b_F$, $d_1$, $d_2$, $d_3$, $d_4$ (see \cite{Borasoy:2006sr,ref3,ref22,ref2,ref4,ref25} for more details).
However only very recently it has started to be used in real calculations and data fitting \cite{Borasoy:2006sr,ref3,ref22,Mai:2012dt,ref2,ref4,ref25,crimea}. The reason is rather straightforward - NLO terms in the chiral Lagrangian depend on 7 new parameters, which were not known, and thus the predictive power of the NLO  U$\chi$PT calculations was rather questionable. 
  
Thanks to great experimental advances of the last years, like for example CLAS photoproduction experiments \cite{CLAS},  we have accumulated a sufficient amount of a good quality data to attempt to fit these new parameters. Also due to the large amount of theoretical studies based on the WT interaction we know where this approach fails to describe the data. In particular, in our study we concentrate on the $\Xi$ hyperon production reactions: $K^- p \rightarrow K^+ \Xi^-\,, K^0 \Xi^0$, where the effect of the NLO terms in the Lagrangian play a crucial role, since the cross sections are zero at tree level of the WT term.  These reactions are also particularly interesting, because they were not considered in the works of the other groups \cite{Borasoy:2006sr,ref3,ref22,Mai:2012dt,ref2,ref4}.

The U$\chi$PT method consists in solving the Lippmann-Schwinger (LS) equations in coupled channels, which is reduced to a system of  algebraic equations \cite{ref1}:
\begin{equation}
T_{ij} =V_{ij}+V_{il} G_l T_{lj}\,,
\label{LS}
\end{equation} 
where $T_{ij}$ is the scattering amplitude for the transition from channel "i" to channel "j"; the subscripts 
run over all the possible channels. In particular, for the meson-baryon interaction in the $S=-1$ sector, which is of prime interest for us,  there are the following 10 channels: $K^- p$, $\bar{K}^0 n$, $\pi^0 \Lambda$, $\pi^0 \Sigma^0$, $\pi^+ \Sigma^-$, $\pi^- \Sigma^+$, $\eta \Lambda$, 
$\eta\Sigma^0$, $K^+ \Xi^-$, $K^0 \Xi^0$. 

In our study we calculate the loop function, $G_l$, using a dimensional regularization scheme: 
\begin{equation}
G_l=\frac{2M_l}{(4\pi)^2}\left\{a_l+\ln\frac{M_l^2}{\mu^2}+\frac{m_l^2-M_l^2+s}{2s}\ln\frac{m_l^2}{M_l^2}+\frac{q_{cm}}{\sqrt{s}}\ln\left[\frac{(s+2\sqrt{s}q_{cm})^2-(M_l^2-m_l^2)^2}{(s-2\sqrt{s}q_{cm})^2-(M_l^2-m_l^2)^2}\right]\right\}\,,
\end{equation}
where $M_l$ and $m_l$ are the baryon and meson masses of the "$l$" channel correspondingly, and $a_l$ are the so called subtraction constants, which are used as free parameters and fitted to the experimental data. Taking into account the isospin symmetry there are only 6 independent subtraction constants. See Ref. \cite{ref1} for more details.

Having calculated the T-matrix by solving the system of equations (\ref{LS}), $T^{LS}(s',s)$,  we can obtain the unpolarized cross-section for the $i \rightarrow j$ reaction in the usual way: 
\begin{equation}
\frac{d\sigma_{ij}}{d\Omega}=\frac{1}{64\pi^2}\frac{4M_iM_j}{s}\frac{k_j}{k_i}M_{ij}\,; \quad M_{ij}=\frac{1}{2}\sum_{s',s}|T^{LS}_{ij}(s',s)|^2
\label{dsigma_0}
\end{equation}

\subsection{Inclusion of heavy high spin hyperonic resonances in $\bar{K} N\rightarrow K\Xi$ transitions}

However, our previous studies \cite{ref25,crimea}, together with the phenomenological work \cite{ref12}, indicate the necessity to take into consideration the  $\bar{K} N \rightarrow Y \rightarrow K \Xi$ reactions, where $Y$ stands for some high spin resonance, which couples to these channels. A lot of $Y$ resonances are known in the relevant energy range. The PDG compilation \cite{PDG} gives eight resonances of four- and three-star status with $1.89<M<2.35$ GeV, see table \ref{tab11}. On the other hand, branching ratios of $K\Xi$ decay are determined for none of them. Only for two of them, upper limits [3\% for $\Lambda(2100)$ and 2\% for $\Sigma(2030)$] are deduced \cite{PDG}.

Naturally, the main decay channels for all these resonances are $\pi\Lambda$ (for $\Sigma$ resonances), $\pi\Sigma$, and $\overline{K}N$, while the branching ratios of $K\Xi$ decay are small, since the $K\Xi$ decay requires creation of an additional $\overline{s}s$ pair and it is not so favorable energetically. On the other hand, cross sections of the $\overline{K}N\to K\Xi$ reaction are more than two orders of magnitude smaller than, for example  $\overline{K}N\to \pi\Sigma$ and $\overline{K}N\to \overline{K}N$,  so even small branching ratios can contribute sizably to the former reaction. Thus, the role of the above-threshold resonances should be studied. Most of these resonances have high spins, and therefore require a special treatment, analogous to that performed in \cite{ref12,ref23,ref24}.

\begin{table}
\begin{tabular}{lllll}
\hline
 {\bf Resonance}   & {\bf $I$ $(J^P)$}   & {\bf Mass $(MeV)$}   &{\bf $\Gamma$ $(MeV)$}  & {\bf $\Gamma_{K\Xi}/\Gamma $}   \\
\hline
 $\Lambda(1890)$ &	$0\left(\frac{3}{2}^+\right)$	& 1850 - 1910	& 60 - 200 & \\
 $\Lambda(2100)$ &	$0\left(\frac{7}{2}^-\right)$ & 2090 - 2110 &	100 - 250 & $< 3\%$ \\
 $\Lambda(2110)$ &	$0\left(\frac{5}{2}^+\right)$ & 2090 - 2140 &	150 - 250 &  \\
 $\Lambda(2350)$ &	$0\left(\frac{9}{2}^+\right)$ & 2340 - 2370 &	100 - 250 &  \\
 $\Sigma(1915)$ &	$1\left(\frac{5}{2}^+\right)$ & 1900 - 1935 &	80 - 160 &  \\
 $\Sigma(1940)$ &	$1\left(\frac{3}{2}^-\right)$ & 1900 - 1950 &	150 - 300 &  \\
 $\Sigma(2030)$ & $1\left(\frac{7}{2}^+\right)$ & 2025 - 2040 & 150 - 200 & $< 2\%$ \\
 $\Sigma(2250)$ &  $1\left(?^?\right)$  & 2210 - 2280 & 60 - 150 &  \\
\hline
\end{tabular}
\caption{Hyperonic  four- and three-star resonances in the range $1.89<M<2.35$ GeV listed in PDG \cite{PDG}. }
\label{tab11}
\end{table}

Based on our fit presented in \cite{crimea} it seems that the $\Sigma(2030)$ and the $\Sigma(2250)$ resonances would be good candidates to be considered, and this observation coincides with the finding of \cite{ref12}.
Spin and parity ($J^\pi =7/2^+$) of the $\Sigma(2030)$ are well established. For the $\Sigma(2250)$, most probable assignments are $5/2^-$ and $9/2^-$ \cite{PDG}. We choose $J^\pi =5/2^-$  to simplify the calculations.\footnote{ It has also been shown in \cite{ref12} that the $9/2^-$ choice do not change the results drastically.} We fix the masses of the resonances to their nominal values, 2030 MeV and 2250 MeV, and their widths to 175 MeV and 105 MeV respectively.
The latter quantities are the median values of the ranges presented for the widths in the PDG compilation \cite{PDG}.

Thus, in order to calculate the  $\overline{K} N \rightarrow K^+ \Xi^-$,  $K^0 \Xi^0$ reaction cross sections we add to the corresponding $T^{LS}(s',s)$ amplitude the contributions from the
$\overline{K} N\rightarrow \Sigma(2030) \rightarrow K\Xi$ and $\overline{K} N\rightarrow \Sigma(2250) \rightarrow K\Xi$ scattering amplitudes, which we denote by  $T^{{7/2}^+}(s',s)$ and $T^{{5/2}^-}(s',s)$ correspondingly. 

As in \cite{ref24}, we adopt the Rarita-Schwinger method to describe spin-5/2 and 7/2 baryon fields, 
which are given as a rank-2 tensor $Y_{5/2}^{\mu\nu}$  and as a rank-3 tensor $Y_{7/2}^{\mu\nu\alpha}$ respectively.

\begin{equation}
\mathcal{L}^{{5/2}^\pm}_{BYK}(q)=i\frac{g_{BY_{5/2}K}}{m_K^2}\bar{B}\Gamma^{(\pm)}Y_{5/2}^{\mu\nu}\partial_\mu\partial_\nu K+H.c.\,, \q \mathcal{L}^{{7/2}^\pm}_{BYK}(q)=-\frac{g_{BY_{7/2}K}}{m_K^3}\bar{B}\Gamma^{(\mp)}Y_{7/2}^{\mu\nu\alpha}\partial_\mu\partial_\nu \partial_\alpha K+H.c.\,,
\label{L_Res}
\end{equation}
where $\Gamma^{(\pm)}= \binom{\gamma_5}{1}$. 
Then the corresponding propagators are given by \cite{ref24}:
\begin{equation}
S_{5/2}(q)=\frac{i}{\slashed{q}-M_{Y_{5/2}}+i\Gamma_{5/2}/2}\Delta^{\beta_1\beta_2}_{\alpha_1\alpha_2}\,, \q  S_{7/2}(q)=\frac{i}{\slashed{q}-M_{Y_{7/2}}+i\Gamma_{7/2}/2}\Delta^{\beta_1\beta_2\beta_3}_{\alpha_1\alpha_2\alpha_3} \,,
\label{S_Res}
\end{equation}
where we have included the decay width, $\Gamma_{J}$, of the corresponding resonance, and tensors $\Delta$ are defined as follows: 
\begin{equation}
\Delta^{\beta_1\beta_2}_{\alpha_1\alpha_2} \left(\frac{5}{2} \right)=\frac{1}{2}\left(\theta^{\beta_1}_{\alpha_1}\theta^{\beta_2}_{\alpha_2}+\theta^{\beta_2}_{\alpha_1}\theta^{\beta_1}_{\alpha_2}\right)-\frac{1}{2}\theta_{\alpha_1\alpha_2}\theta^{\beta_1\beta_2}-\frac{1}{10}\left(\bar{\gamma}_{\alpha_1}\bar{\gamma}^{\beta_1}\theta^{\beta_2}_{\alpha_2}+\bar{\gamma}_{\alpha_1}\bar{\gamma}^{\beta_2}\theta^{\beta_1}_{\alpha_2}+\bar{\gamma}_{\alpha_2}\bar{\gamma}^{\beta_1}\theta^{\beta_2}_{\alpha_1}+\bar{\gamma}_{\alpha_2}\bar{\gamma}^{\beta_2}\theta^{\beta_1}_{\alpha_1}\right)\,,
\label{Delta_5}
\end{equation} 
\begin{equation}
\Delta^{\beta_1\beta_2\beta_3}_{\alpha_1\alpha_2\alpha_3} \left(\frac{7}{2} \right)=\frac{1}{36}\sum_{P(\alpha)P(\beta)}\left(\theta^{\beta_1}_{\alpha_1}\theta^{\beta_2}_{\alpha_2}\theta^{\beta_3}_{\alpha_3}-\frac{3}{7}\theta^{\beta_1}_{\alpha_1}\theta_{\alpha_2\alpha_3}\theta^{\beta_2\beta_3}-\frac{3}{7}\bar{\gamma}_{\alpha_1}\bar{\gamma}^{\beta_1}\theta^{\beta_2}_{\alpha_2}\theta^{\beta_3}_{\alpha_3}+\frac{3}{35}\bar{\gamma}_{\alpha_1}\bar{\gamma}^{\beta_1}\theta_{\alpha_2\alpha_3}\theta^{\beta_2\beta_3}\right)\,,
\label{Delta_7}
\end{equation}
where $\theta^{\nu}_{\mu}=g^{\nu}_{\mu}-\frac{q_\mu q^\nu}{M_Y^2}$ , $\bar{\gamma}_{\mu}=\gamma_{\mu}-\frac{q_\mu \slashed{q}}{M_Y^2}$ with $M_Y$ being the pertinent resonance mass. In eq. (\ref{Delta_7}) the tensor for the spin-7/2 field contains summation over all possible permutations of Dirac indexes $\{\alpha_1 \alpha_2 \alpha_3\}$ and $\{\beta_1 \beta_2 \beta_3\}$.

From the Lagrangians of eq. (\ref{L_Res}) one derives the baryon-kaon-$Y_{J}$ vertices: 
\begin{equation}
v^{{5/2}^\pm}_{BYK}=i\frac{g_{BY_{5/2}K}}{m_K^2}k_\mu k_\nu \Gamma^{(\pm)}\,, \q v^{{7/2}^\pm}_{BYK}=-\frac{g_{BY_{7/2}K}}{m_K^3}k_\mu k_\nu k_\sigma \Gamma^{(\mp)}\,,
\label{Ver_Res}
\end{equation}
and the scattering amplitude for  $\overline{K} N\rightarrow K\Xi$ reaction can be obtained straightforwardly:
\begin{equation}
T^{{5/2}^-}(s',s) = \frac{g_{ \Xi Y_{5/2}K} g_{NY_{5/2}\overline{K}}}{m_K^4}\bar{u}_\Xi^{s'}(p')\frac{k'_{\beta_1}k'_{\beta_2}\Delta^{\beta_1\beta_2}_{\alpha_1\alpha_2}k^{\alpha_1}k^{\alpha_2}}{\slashed{q}-M_{Y_{5/2}}+i\Gamma_{5/2}/2}u_N^{s}(p)\exp\left(-\vec{k}^2/\Lambda^2_{5/2}\right)\exp\left(-\vec{k'}^2/\Lambda^2_{5/2}\right)\,,
\label{T_5}
\end{equation}
\begin{equation}
T^{{7/2}^+}(s',s) = \frac{g_{ \Xi Y_{7/2}K} g_{NY_{7/2}\overline{K}}}{m_K^6}\bar{u}_\Xi^{s'}(p')\frac{k'_{\beta_1}k'_{\beta_2}k'_{\beta_2}\Delta^{\beta_1\beta_2\beta_3}_{\alpha_1\alpha_2\alpha_3}k^{\alpha_1}k^{\alpha_2}k^{\alpha_3}}{\slashed{q}-M_{Y_{7/2}}+i\Gamma_{7/2}/2} u_N^{s}(p)\exp\left(-\vec{k}^2/\Lambda^2_{7/2}\right)\exp\left(-\vec{k'}^2/\Lambda^2_{7/2}\right)\,,
\label{T_7}
\end{equation}
where we have introduced phenomenological form factors, $\exp\left(-\vec{q\,}^2/\Lambda^2_{J}\right)$, associated with each vertex, in order to suppress high momentum contributions, as it was also done in \cite{ref12}.

\begin{figure}
  \includegraphics[width=.67\textwidth]{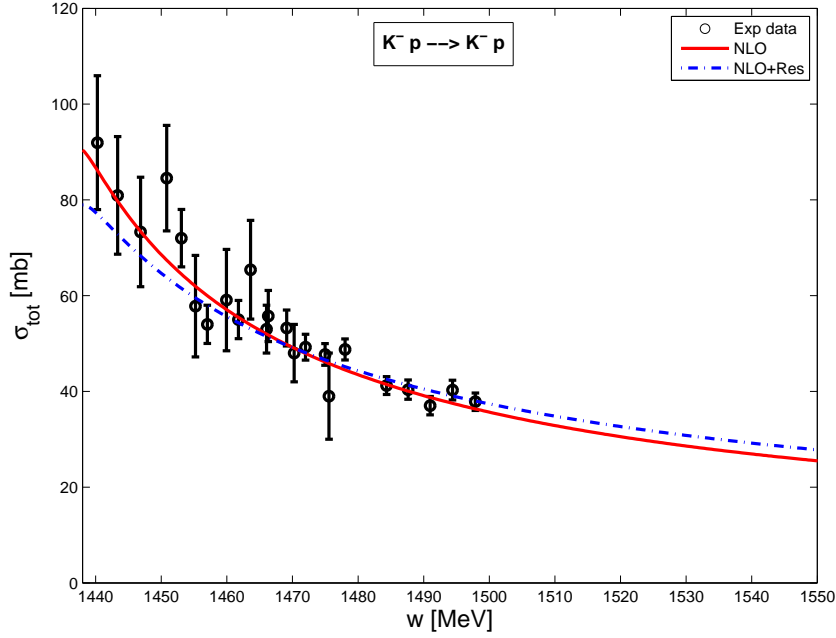}
\caption{The total cross section of the $K^- p\to K^- p$ reaction.  The solid line represents results of NLO fit, the dashed-dotted line - NLO+Res fit, see text for more details. Experimental data are from \cite{ref13,ref14,ref15,ref16,ref17,ref18,ref19}. } 
  \label{fig1}
\end{figure}

Finally, for the initial channels $i=K^- p$, $\overline{K}^0 n$ and final channels $j=K^+ \Xi^-$, $K^0 \Xi^0$ we obtain
\begin{equation}
T^{tot}_{ij}(s',s)= \sqrt{4M_pM_\Xi} T^{LS}_{ij}(s',s)+T^{{5/2}^-}(s',s)+T^{{7/2}^+}(s',s)\,.
\end{equation}
And then we can proceed following Eqs. (\ref{dsigma_0}).

\begin{figure}
  \includegraphics[width=.67\textwidth]{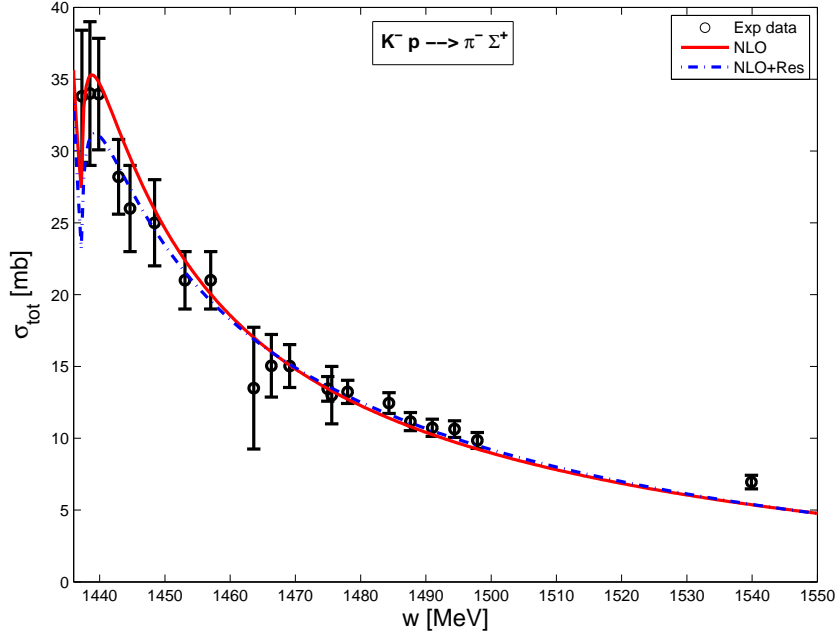}
\caption{The total cross section of the $K^- p\to \pi^- \Sigma^+$ reaction.  The solid line represents results of NLO fit, the dashed-dotted line - NLO+Res fit, see text for more details. Experimental data are from \cite{ref13,ref14,ref15,ref16,ref17,ref18,ref19}.} 
  \label{fig2}
\end{figure}

\section{Results and conclusions}

\begin{figure}
  \includegraphics[width=.67\textwidth]{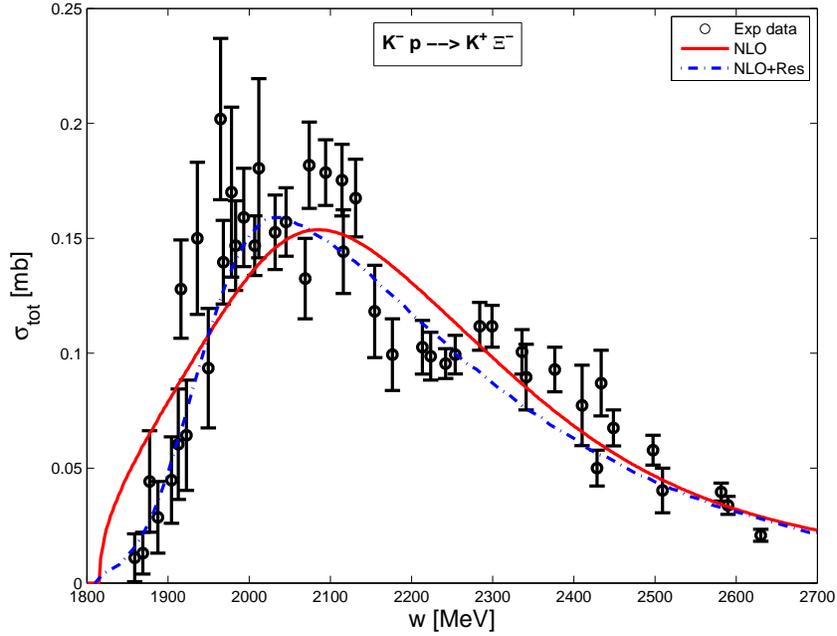}
\caption{The total cross section of the $K^- p\to K^+ \Xi^-$ reaction.  The solid line represents results of NLO fit, the dashed-dotted line - NLO+Res fit, see text for more details. Experimental data are from \cite{ref13,ref14,ref15,ref16,ref17,ref18,ref19}.} 
  \label{fig3}
\end{figure}

\begin{figure}
  \includegraphics[width=.67\textwidth]{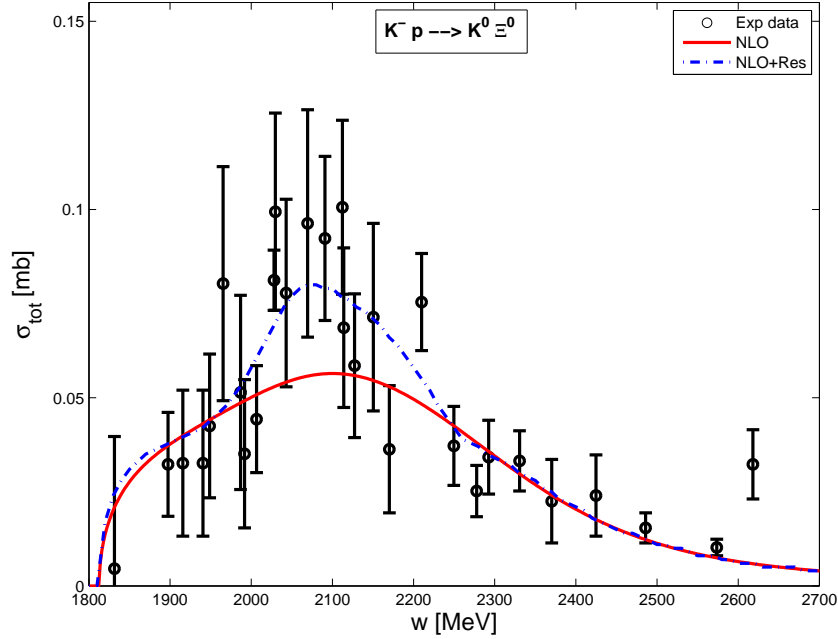}
\caption{The total cross section of the $K^- p\to K^0 \Xi^0$ reaction.  The solid line represents results of NLO fit, the dashed-dotted line - NLO+Res fit, see text for more details. Experimental data are from \cite{ref13,ref14,ref15,ref16,ref17,ref18,ref19}.} 
  \label{fig4}
\end{figure}

We will fit the available experimental data on cross sections for the transitions from the $K^- p$ initial state to different final states, some examples are presented in Figs. \ref{fig1}-\ref{fig4}, as well as the threshold branching ratios:
$$\gamma=\frac{\Gamma(K^- p \rightarrow \pi^+ \Sigma^-)}{\Gamma(K^- p \rightarrow \pi^- \Sigma^+)},  \  R_n=\frac{\Gamma(K^- p \rightarrow \pi^0 \Lambda)}{\Gamma(K^- p \rightarrow neutral\, states)},  \
R_c=\frac{\Gamma(K^- p \rightarrow \pi^+ \Sigma^-,\pi^- \Sigma^+ )}{\Gamma(K^- p \rightarrow inelastic\, channels)}\,,$$ 
given in Table \ref{tab1}.

\begin{table}
\begin{tabular}{lccc}
\hline
  & \tablehead{1}{c}{b}{$\gamma$}
  & \tablehead{1}{c}{b}{$R_n$}
  & \tablehead{1}{c}{b}{$R_c$}   \\
\hline
NLO &	2.36 &	0.197 &	0.659  \\%
NLO+RES & 2.37 & 0.193 & 0.657  \\%
Exp. &	$2.36\pm 0.04$ &	$0.189\pm 0.015$ & $0.664\pm 0.011$ \\
\hline
\end{tabular}
\caption{This table shows the branching ratios at threshold for the  NLO and NLO+Res fits, presented in Table \ref{tab_a}, and the corresponding experimental values.}
\label{tab1}
\end{table}

We perform 14 and 17 parameter fits:\\ 
1) NLO interaction kernel without additional resonance contributions (NLO fit), 14 free parameters: the pion decay constant, 6 subtraction constants, and 7 NLO parameters ($b_0$, $b_D$, $b_F$, $d_1$, $d_2$, $d_3$, $d_4$);\\
2) NLO interaction kernel with high spin resonance contributions in $K^- p \to K^+ \Xi^-$, $K^0 \Xi^0$ channels (NLO+Res fit), 17 free parameters: 14 are the same as above, the new parameters are: effective couplings of the $Y_{5/2}$ and $Y_{7/2}$ resonances, and $\Lambda_{J}$ parameter which defines the form factors of the resonances (as a first approximation we choose $\Lambda_{5/2}=\Lambda_{7/2}$). \\
The obtained parameter values and resulting $\chi^2/d.o.f.$ are presented in Table \ref{tab_a}.

First of all, we would like to stress that the inclusion of the NLO terms into the interaction kernel substantially improves the agreement with data with respect to WT fits. This has been shown in detail in Refs. \cite{ref2,ref3,ref4,ref25,ref22}, and we are not going to discuss these aspects in the present letter.


\begin{table}
\begin{tabular}{lrr}
\hline
  & \tablehead{1}{r}{b}{NLO}
  & \tablehead{1}{r}{b}{NLO+RES} \\
\hline
$a_{\overline{K}N} \ (10^{-3})$ & 4.64 & 6.84     \\
$a_{\pi\Lambda}\ (10^{-3})$ & 24.52 & 28.75  \\
$a_{\pi\Sigma}\ (10^{-3})$ & 2.06 & 0.79  \\
$a_{\eta\Lambda}\ (10^{-3})$ & -9.10 & -11.53  \\
$a_{\eta\Sigma}\ (10^{-3})$ & -8.43 & -13.53  \\
$a_{K\Xi}\ (10^{-3})$ & 37.54 & 29.71  \\
$f $  & 1.19$f_\pi$ & 1.18$f_\pi$  \\
$b_0 \ (GeV^{-1}) $ & -0.37 & -0.31   \\
$b_D \ (GeV^{-1}) $ & -0.04 & - 0.11   \\
$b_F \ (GeV^{-1}) $ & 0.37 & 0.26 \\
$d_1 \ (GeV^{-1}) $ & 0.24 & 0.14   \\
$d_2 \ (GeV^{-1}) $ & 0.41 & 0.42   \\
$d_3 \ (GeV^{-1}) $ & 0.74 & 0.74   \\
$d_4 \ (GeV^{-1}) $ & -0.58 & -0.60  \\
$g_{ \Xi Y_{5/2}K}\cdot g_{NY_{5/2}\overline{K}}$ & - & 84.12 \\
$g_{ \Xi Y_{7/2}K}\cdot g_{NY_{7/2}\overline{K}}$ & - & 283.47 \\
$\Lambda_{5/2}=\Lambda_{7/2}\ (MeV) $ & - & 308.13 \\
\hline
$\chi^2/d.o.f.$ &1.88 & 1.40 \\
\hline
\end{tabular}
\caption{Results of the NLO (14 parameters) and NLO+Res (17 parameters) fits, see text for more details. }
\label{tab_a}
\end{table}

Looking at Figs. \ref{fig1}, \ref{fig2} and Tables \ref{tab1}, \ref{tab_a} we can conclude that for the observables unrelated to the $\Xi$ hyperon production reactions the model is only slightly modified, although the agreement with data is a little bit better. While on Figs. \ref{fig3}, \ref{fig4} we see a substantial improvement in describing the $\bar{K} N\rightarrow K\Xi$ transitions, which are the key point of this work, as well as of the previous study \cite{ref25,crimea}.  The improvement in these two channels is the main reason of the $\chi^2/d.o.f.$ reduction from NLO fit to NLO+Res one, see Table \ref{tab_a}. 

As it was discussed above these channels are extremely sensitive to the NLO corrections and therefore may play a crucial role in determining the NLO parameters. Comparing the NLO and NLO+Res fits in Table \ref{tab_a} we can see that the most drastic change has to do with the $b_D$ parameter, the other sensitive ones are  $b_F$, $d_1$, and $b_0$.

The final goal of our study is to find trustable restrictions on the 7 NLO parameters of the chiral Lagrangian. We would like to stress that, technically, to change in the calculations the WT interaction, eq. (\ref{WT}), to the NLO interaction, eq. (\ref{V_NLO}), is rather straightforward. The problem comes from the fact that the 7 new parameters of the NLO interaction are not well controlled at the moment. Once stable values 
for these parameters are obtained, all the groups doing simulations based on the chiral Lagrangian will be able to increase the accuracy of their calculations to the next order with a rather little effort. Based on the presented results we believe that the inclusion of high spin hyperonic resonances in $K^- p \to K^+ \Xi^-$, $K^0 \Xi^0$ channels (which are 0 at three level in LO) is mandatory in order to obtain reliable NLO parameters.


\begin{theacknowledgments}
This work is supported  by the European Community - Research Infrastructure Integrating Activity Study of Strongly Interacting Matter (HadronPhysics3, Grant Agreement Nr 283286) under the 7th Framework Programme, by the contract FIS2011-24154 from MICINN (Spain), and by the Ge\-ne\-ra\-li\-tat de Catalunya, contract 2009SGR-1289.
\end{theacknowledgments}



\bibliographystyle{aipproc}   


\end{document}